\documentclass[10pt,a4paper]{article}

\newif\ifconf
\conffalse

\newcommand{\newparagraph}[1]{\paragraph{#1}}
\usepackage[affil-it]{authblk}
\author[1]{Ma\"el Dumas} 
\author[1]{Anthony Perez}
\author[1]{Ioan Todinca}
\affil[1]{Univ. Orl\'eans, INSA Centre Val de Loire, LIFO EA 4022, F-45067 Orl\'eans, France}
\usepackage{version}
\usepackage{fullpage}

\def\ie{{\em i.e.}~}

\def\TPC{{\textsc{Trivially Perfect Completion}}}
\def\TPD{{\textsc{Trivially Perfect Deletion}}}
\def\TPE{{\textsc{Trivially Perfect Editing}}}

\newcommand{\T}{\mathcal{T}}
\newcommand{\B}{\mathcal{B}}

\newcommand{\F}{\mathcal{F}}

\title{A cubic vertex-kernel for \TPE{}}

\usepackage{enumitem}

\usepackage[normalem]{ ulem }
\usepackage{soul}

\usepackage[numbers,sort]{natbib}
\usepackage[svgnames,x11names]{xcolor}

\usepackage[colorinlistoftodos,backgroundcolor=Yellow1!80,linecolor=OrangeRed1]{todonotes}
\usepackage{subfigure}

\usepackage[utf8]{inputenc}
 
\usepackage{fixmath}
\usepackage{stmaryrd}
\usepackage{graphicx}

\usepackage{amsmath, amssymb, amsthm}

\newtheorem{lemma}{Lemma}

\newtheorem{proposition}{Proposition}
\newtheorem{definition}{Definition}
\theoremstyle{definition}

\newtheorem{polyrule}{Rule}

\usepackage[linesnumbered,vlined,ruled]{algorithm2e}

\newcommand{\Pb}[4]{%
\begin{center}
  \begin{tabular}{|l|}%
  \hline
    \begin{minipage}[c]{0.95\linewidth}
      \smallskip%
      \par\noindent%
      \centerline{\underline{\textsc{#1}}}%
      \par\noindent%
      \textbf{\textsf{Input}}: #2% 
      \par\noindent%
      \textbf{\textsf{#4}}: Does there exist %
      #3?% 
      \smallskip%
      \par\noindent%
    \end{minipage}
  \\\hline
  \end{tabular}%
\end{center}
}%

\usepackage{proof-at-the-end}
\pgfkeys{/prAtEnd/global custom defaults/.style={%
normal,%
no link to proof
}}

\NewDocumentEnvironment{proofApx}{O{}+b}{
  \pgfkeys{%
    /prAtEnd/.cd,
    all defaults,
    prAtEndTmpStyle/.style/.expand once={\pratendlastoptions},
    prAtEndTmpStyle,
    #1
  }%
  \unless\ifallattheend
    \iflinktoproof%
      \pratendtextlink{}%
    \fi%
    \ifproofhere%
      \begin{proof}%
        #2%
      \end{proof}%
    \fi%
  \fi%
  \ifproofend%
    \appendtofile{\prefixPrAtEndFiles\category.tex}{\string\label{proofsection:prAtEnd\pratendcountercurrent}\string\begin{proof}[\pratendtextproof]\string\phantomsection\string\label{proof:prAtEnd\pratendcountercurrent}\detokenize{#2}\string\end{proof}}%
  \fi%
}{}

\NewDocumentEnvironment{proofApxClaim}{O{}+b}{
  \pgfkeys{%
    /prAtEnd/.cd,
    all defaults,
    prAtEndTmpStyle/.style/.expand once={\pratendlastoptions},
    prAtEndTmpStyle,
    #1
  }%
  \unless\ifallattheend
    \iflinktoproof%
      \pratendtextlink{}%
    \fi%
    \ifproofhere%
      \begin{proof}%
        #2%
      \end{proof}%
    \fi%
  \fi%
  \ifproofend%
    \appendtofile{\prefixPrAtEndFiles\category.tex}{\string\label{proofsection:prAtEnd\pratendcountercurrent}\string\begin{proof}[\pratendtextproof]\string\phantomsection\string\label{proof:prAtEnd\pratendcountercurrent}\detokenize{#2}
    \string\string\hfill\string\qed\string\end{proof}}%
  \fi%
}{}

\begin{document}

\maketitle

\begin{abstract}
    We consider the \TPE{} problem, where one is given an undirected graph $G = (V,E)$ and a parameter 
    $k \in \mathbb{N}$ and seeks to \emph{edit} (add or delete) at most $k$ edges from $G$ to obtain 
    a trivially perfect graph. The related \TPC{} and \TPD{} problems are obtained by only allowing edge 
    additions or edge deletions, respectively. Trivially perfect graphs are both chordal and cographs, 
    and have applications related to the tree-depth width parameter and to social network analysis. All variants of the problem are known to be NP-Complete~\cite{BBD06,NG13} and to admit so-called polynomial kernels~\cite{DP18,Guo07}. 
    More precisely, the existence 
    of an $O(k^3)$ vertex-kernel for \TPC{} was announced by Guo~\cite{Guo07} but without a stand-alone proof. 
    More recently, Drange and Pilipczuk~\cite{DP18} provided $O(k^7)$ vertex-kernels for these  problems 
    and left open the existence of cubic vertex-kernels. 
    In this work, we answer positively to this question for all three variants of the problem. 
\end{abstract}

\section*{Introduction}

A broad range of optimization problems on graphs are particular cases of so-called modification problems. Given an arbitrary graph $G=(V,E)$ and an integer $k$, the question is whether $G$ can be turned into a graph satisfying some desired property by at most $k$ \emph{modifications}. By modifications we mean, according to the problem, vertex deletions (as for \textsc{Vertex Cover} and \textsc{Feedback Vertex Set} where we aim to obtain graphs with no edges, or without cycles respectively) or edge deletions and/or additions (as for \textsc{Minimum Fill-In}, also known as \textsc{Chordal Completion}, where the goal is to obtain a chordal graph, with no induced cycles with four or more vertices, by adding at most $k$ edges). 

Here we consider edge modifications problems, that can be split in three categories, depending whether we allow only edge additions, only edge deletions, or both operations, in which case we speak of edge editing. Consider a family $\mathcal{H}$ of graphs, called \emph{obstructions}. In the {\sc $\mathcal{H}$-free editing} problem we seek to 
edit at most $k$ edges of $G$ to obtain a graph that does not contain 
any obstruction from $\mathcal{H}$ as an induced subgraph. 
One can similarly define {\sc $\mathcal{H}$-free completion} 
and {\sc $\mathcal{H}$-free deletion} variants of this problem by only allowing the addition or deletion of edges, 
respectively. 
E.g., \textsc{Minimum Fill-In} corresponds to \textsc{$\mathcal{H}$-free completion}, where $\mathcal{H}$ is formed by all cycles with at least four vertices. 
For most families $\mathcal{H}$, all three versions are NP-complete, but thinking of $k$ as of some suitably small quantity, they have been intensively studied in the framework of 
\emph{parameterized complexity} (see~\cite{CGF+20} for a comprehensive survey).
The aim of parameterized complexity is to determine whether it is possible to decide 
the instance at hand 
in time $f(k) \cdot n^{O(1)}$ for some computable function $f$. Such problems are said to be FPT (\emph{fixed-parameter tractable}). 
With a simple but elegant and powerful argument, Cai~\cite{Cai96} proved that whenever $\mathcal{H}$ is finite all three variants are FPT. Basically, whenever the graph contains one of the obstructions (graphs of $\mathcal{H}$), the algorithm branches on all possible modifications to destroy it, and makes the recursive calls with a lesser parameter $k$.
When the family $\mathcal{H}$ contains all cycles with at least four vertices, the corresponding edition problem {\sc Chordal Editing}  
 was shown to be FPT relatively recently~\cite{CM16}. The completion variant, i.e., the 
{\sc Minimum Fill-in}, was known to be FPT since the 90's~\cite{Cai96,KST99}. \\

We consider an equivalent definition of fixed-parameter tractability, namely \emph{kernelization}. Given 
a parameterized problem $\Pi$, a \emph{kernelization algorithm} for $\Pi$ 
(or \emph{kernel} for short) is an 
algorithm that given any instance $(I,k)$ of $\Pi$ runs in time polynomial in $|I|$ and $k$ and outputs an equivalent 
instance $(I',k')$ of $\Pi$ such that $|I'| \leqslant h(k)$ and $k' \leqslant g(k)$ for some computable 
functions $g$ and $h$. Whenever $h$ is polynomial, we say that $\Pi$ admits a \emph{polynomial} kernel. 
A kernelization algorithm uses a set of polynomial-time computable \emph{reduction rules} to \emph{reduce} the instance at hand. We say that a reduction rule is \emph{safe} whenever its application 
on an instance $(I,k)$ of $\Pi$ results in an equivalent instance $(I',k')$ of $\Pi$. 
It is well-known that a parameterized problem is FPT if and only if it admits a kernelization algorithm~\cite{FLS+19}. 
While many polynomial kernels are known to exist for editing problems (see~\cite{CGF+20} or~\cite{LWG14} for  surveys), 
it is known that some editing problems 
are unlikely to admit polynomial kernels under reasonable theoretical complexity assumptions~\cite{CC15,GHP+13,KW09}. 
When $\mathcal{H}$ contains only a single obstruction, several results   
towards a dichotomy regarding the existence of polynomial kernels 
have been obtained~\cite{ASS17,CC15,MS20}. Very recently, Marx and Sandeep~\cite{MS20} 
narrowed down the problem for obstructions containing at least $5$ vertices to only nine distinct obstructions. In other words, the non-existence of polynomial kernels for {\sc $\mathcal{H}$-free editing}  
for all such obstructions
would imply the non-existence of polynomial kernels for any obstruction with at least $5$ vertices. 
When $\mathcal{H}$ contains several obstructions, a very natural setting is to include all cycles 
in $\mathcal{H}$, thus targeting a subclass of chordal graphs. Indeed, editing (and especially completion) problems towards such classes cover classical problems with both theoretical and 
practical interest~\cite{EEC88,GKS94,HSS01,KST99,Yannakakis81}.  
Notice that many known polynomial kernels for editing problems concern such classes~\cite{BPP10,BP13,DP18,Guo07,KST99}. For completion and deletion versions, polynomial kernels are often used
as a first
step in the design of subexponential parameterized algorithms~\cite{BFP+15,DFP+15,FV13,GKK+15}. \\ 

In this work, we focus on editing problems towards trivially perfect graphs, that is $\mathcal{H} = \{P_4, C_4\}$ 
(respectively a path and a cycle on $4$ vertices). This problem is known as \TPE{} in the literature. By allowing edge addition or edge deletion only, we obtain the \TPC{} and \TPD{} problems, respectively. 

\newparagraph{Related work.} While the NP-Completeness of \TPC{} and \TPD{} has been known for some time~\cite{BBD06}, the complexity of \TPE{} remained open until a work of Nastos and Gao~\cite{NG13}. 
Trivially perfect graphs have recently regained attention since they are related to the well-studied width parameter \emph{tree-depth}~\cite{Golumbic78,NdM15} which corresponds to the size of the largest clique of a trivially perfect supergraph of $G$ with the smallest clique number. 
Moreover, Nastos and Gao~\cite{NG13} proposed a new definition for community structure based on small obstructions. 
In particular, the authors emphasized that editing a given graph into a trivially perfect graph 
\emph{yields meaningful clusterings in real networks}~\cite{NG13}. 
Trivially perfect graphs also correspond to chordal cographs and admit 
a so-called \emph{universal clique decomposition}~\cite{DFP+15}.  
Polynomial kernels with $O(k^7)$ vertices 
have been obtained for all variants of the problem by Drange and Pilipczuk~\cite{DP18}. 
The technique used 
relies on a reduction rule bounding the number of vertices in any trivially perfect \emph{module} and the 
computation of a so-called \emph{vertex modulator}, that is a maximal packing of obstructions with additional properties. 
Combined with sunflower-like reduction rules and a careful analysis of the graph remaining apart from the vertex 
modulator, the authors managed to provide polynomial kernels. They then asked whether the $O(k^7)$ bound could be 
improved, and qualify as ``really challenging question'' whether one can match the $O(k^3)$ bound for \TPC{} claimed by Guo~\cite{Guo07}.

\newparagraph{Our contribution.} We answer positively to this question and provide kernels with 
$O(k^3)$ vertices for all considered problems. To be complete, a quadratic kernel for the completion version only is claimed in~\cite{BBP21,CK21}. While our kernelization algorithm shares similarities with the work of Drange and Pilipczuk~\cite{DP18}, our technique differs in several points. In particular, we do not rely on 
the computation of a vertex modulator, a useful technique to design polynomial kernels but somehow responsible for the large bound obtained. 
To circumvent this issue, we only rely on the so-called universal clique decomposition of trivially perfect graphs. This decomposition  partitions the vertices of trivially perfect graph $G$ into cliques, the bags being structured as nodes of a rooted forest such that two vertices are adjacent in $G$ if and only they are in a same bag, or in two bags such that one is an ancestor of the other in the forest. 
For any positive instance of the problem, at most $2k$ bags contain vertices incident to modified edges. The rest of the bags can be regrouped into two type of chunks. Some correspond to trivially perfect modules of the input graph (which are known to be reducible to small sizes by~\cite{DP18}, as well as the bags~\cite{BPP10}), other have a more complicated but still particular structure, similar to the \emph{combs} of~\cite{DP18}. 
We show how to reduce the size of these combs. Altogether we believe that our rules not only improve the size of the kernel but also significantly simplify the kernelization algorithm of~\cite{DP18}. 
Last but not least, we think that this
approach based on tree-like decompositions and the analysis of large chunks of the graph that are not affected by the modified edges might be exploitable for other editing problems. Indeed the technique has strong similarities with the notion of 
branches introduced by Bessy et al.~\cite{BPP10} for modification to $3$-leaf power graphs, a closely related graph class. 

\newparagraph{Outline.} We begin with some preliminaries definitions and results about trivially perfect graphs 
(Section~\ref{sec:prelim}). We then introduce the notion of combs and provide the set of reduction rules needed to obtain an $O(k^3)$ vertex-kernel 
for \TPE{} (Section~\ref{sec:rules}). 
Details of the kernelization algorithm, especially on finding large combs, are given in Section~\ref{sec:haircut}, and 
the combinatorial bound on the kernel size is provided in Section~\ref{sec:size}. 
We explain how these results can be adapted to obtain similar kernels for \TPC{} and \TPD{} in Section~\ref{sec:variants}. The Conclusion section summarizes the results and suggests further developments. 

\section{Preliminaries}
\label{sec:prelim}

We consider simple, undirected graphs $G = (V,E)$ where $V$ denotes the \emph{vertex set} and $E \subseteq (V \times V)$ the % 
\emph{edge set} of $G$. We will sometimes use $V(G)$ and $E(G)$ to clarify the context. 
Given a vertex $u \in V$, the \emph{open neighborhood} of $u$ is the set % 
$N_G(u) = \{v \in V:\ uv \in E\}$. 
The \emph{closed neighborhood} of $u$ is defined as $N_G[u] = N_G(u) \cup \{u\}$. 
A vertex $u \in V$ is \emph{universal} if $N_G[u] = V$, and 
two vertices $u$ and $v$ are \emph{true twins} if $N_G[u] = N_G[v]$. 
Given a subset of vertices $S \subseteq V$, $N_G[S]$ is the set $\cup_{v \in S} N_G[v] $
and $N_G(S)$ is the set $N_G[S] \setminus S$. 
We will omit the mention to $G$ whenever the context is clear. 
The subgraph \emph{induced} by $S$ is defined as % 
$G[S] = (S,E_S)$ where $E_S = \{uv \in E:\ u \in S, v \in S\}$. 
For the sake of readability, given a subset $S \subseteq V$ 
we define $G\setminus S$ as $G[V\setminus S]$. % 
A subset of vertices $C \subseteq V$ is a \emph{connected component} of $G$ if $G[C]$ is a maximal connected subgraph of $G$. 
A subset of vertices $M \subseteq V$ is a \emph{module} of $G$ iff $N_G(u) \setminus M = N_G(v) \setminus M$ holds for every $u,v \in M$. 
A maximal set of true twins $K \subseteq V$ is a \emph{critical clique}. 
Notice that $G[K]$ is a clique module and that the set $\mathcal{K}(G)$ of critical cliques of any graph $G$ partitions its vertex set $V(G)$.  

\begin{figure}[ht]
    \centering
    \includegraphics[scale=2.25]{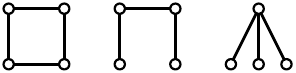}
    \caption{The $C_4$, $P_4$ and claw graphs, respectively. The claw will be useful in some of our proofs.}
    \label{fig:obs}
\end{figure}

\newparagraph{Trivially perfect graphs.} A graph $G = (V,E)$ is trivially perfect if and only 
if it does not contain any $P_4$ (a path on $4$ vertices) nor $C_4$ (a cycle on $4$ vertices) as an induced subgraph (see Figure~\ref{fig:obs}). 
We consider the following problem. 

\Pb{Trivially Perfect Editing}%
{A graph $G=(V,E)$, a parameter $k \in \mathbb{N}$}%
{a set of pairs $F \subseteq (V \times V)$ of size at most $k$ such that the graph 
$H = (V, E \triangle F)$ is trivially perfect, with $E \triangle F = (E \setminus F) \cup (F \setminus E)$}
{Question}

Given an instance $(G = (V,E), k)$ of \TPE{}, a set $F \subseteq (V \times V)$ such that  
$H = (V, E \triangle F)$ is trivially perfect is an \emph{edition} of $G$. 
When $F$ is constrained to be disjoint from (resp. contained in) $E$, we say that $F$ is a \emph{completion} (resp. a \emph{deletion}) 
of $G$. The corresponding problems are \TPC{} and \TPD{}, respectively. 
For the sake of 
simplicity, given an edition (resp. completion, deletion) $F$ of $G$, we use $G \triangle F$, $G+F$ and $G-F$ to 
denote the trivially perfect graphs $(V,E\triangle F)$, $(V, E \cup F)$ and $(V, E \setminus F)$, respectively. A vertex is \emph{affected} by $F$ whenever it is contained in some pair of $F$. The set $F$ is a \emph{$k$-edition} (resp. \emph{$k$-completion}, \emph{$k$-deletion}) whenever $|F| \leqslant k$. Finally, we say that such a set $F$ is \emph{optimal} whenever it is minimum-sized. 

Trivially perfect graphs are 
hereditary and closed under true twin addition. This property will be useful to deal with 
critical cliques, as stated by the following result. Recall that critical cliques are maximal sets of true twins (or, equivalently, maximal clique modules), they will play a central role throughout this paper. 

\begin{lemma}[\cite{BPP10}]
    \label{lem:homogen-compl-here} 
    Let $\mathcal{G}$ be an hereditary class of graphs closed under true twin addition. % 
    For every graph $G=(V,E)$, there exists an optimal edition (resp. completion, deletion) 
    $F$ into a graph of $\mathcal{G}$ such that for any two %
    critical cliques $K$ and $K$' either $(K \times K') \subseteq F$ or % 
    $(K \times K') \cap F = \emptyset$.
\end{lemma}

Several characterizations 
are known to exist for trivially perfect graphs. We will mainly use the following ones.

\begin{proposition}[\cite{YCC96}]
The class of trivially perfect graphs can be defined recursively as follows:
\begin{itemize}
    \item a single vertex is a trivially perfect graph.
    \item Adding a universal vertex to a trivially perfect graph results in a trivially perfect graph.
    \item The disjoint union of two trivially perfect graphs results in a trivially perfect graph.
\end{itemize}

\end{proposition}

\begin{definition}[Universal clique decomposition,~\cite{DFP+15}]
\label{def:ucd}
A universal clique decomposition (UCD) of a connected graph $G=(V,E)$ is a pair $\T = (T= (V_T,E_T), \B = \{B_t\}_{t\in V_T})$ where $T$ is a rooted tree and $\B$ is a partition of the vertex set $V$ into disjoint nonempty subsets, such that:
\begin{itemize}
    \item if $vw \in E$ and $v \in B_t$, $w \in B_s$ then $s$ and $t$ are on a path 
    from a leaf to the root, with possibly $s=t$, and 
    \item for every node $t \in V_T$, the set of vertices $B_t$ is the universal clique of the induced subgraph $G[\bigcup_{s\in V(T_t)}B_s]$, where $T_t$ denotes the subtree of $T$ rooted at $t$.
\end{itemize}
\end{definition}

The vertices of $T$ are called \emph{nodes} of the decomposition, while the sets 
of $\mathcal{B}$ are called \emph{bags}. We will sometimes abuse notation and identify nodes 
of $T$ with their corresponding bags in $\mathcal{B}$. Notice moreover that in a 
universal clique decomposition, every node $t$ of $T$ that is not a leaf has at least 
two children since otherwise $B_t$ would not contain \emph{all} universal vertices of 
$G[\bigcup_{s\in V(T_t)}B_s]$.

\begin{lemma}[\cite{DFP+15}]
\label{lem:ucd}
A connected graph $G$ admits a universal clique decomposition if and only if it is trivially perfect. Moreover, such a decomposition is unique up to isomorphisms.
\end{lemma}

One can observe that finding a universal clique decomposition can be done in polynomial   
time by iteratively identifying universal cliques and connected components. 
Finally, both Definition~\ref{def:ucd} and Lemma~\ref{lem:ucd} can be naturally 
extended to \emph{disconnected} trivially perfect graphs by considering a 
\emph{rooted forest} instead of a rooted tree. More precisely, the universal clique 
decomposition of a disconnected graph $G = (V,E)$ is a rooted forest of universal 
clique decompositions of its connected components. Such a graph is thus trivially 
perfect if and only if it admits a universal clique decomposition shaped like 
a rooted forest. \\

We conclude this section by providing a new characterization of trivially perfect graphs 
in terms of maximal cliques and nested families. 

\begin{definition}[Nested family]
Let $U$ be a universe and $\F\subseteq 2^{|U|}$ a family of subsets of $U$. The family $\F$ is nested iff for every $A,B\in \F$, $A \subseteq B$ or $B\subseteq A$ holds.
\end{definition} 

\begin{theoremEnd}[category=preliminaries]{lemma}
\label{thm:TP_recons}
Let $G=(V,E)$ be a graph, $S \subseteq V $ a maximal clique of $G$ and $K_1,...,K_r$ the connected components of $G\backslash S$. The graph $G$ is trivially perfect if and only if the following conditions are verified\: 
\begin{enumerate}[label=(\roman*)] 
    \item\label{it1:recons}  $G[S \cup K_i]$ is trivially perfect for every $1\leq i \leq r$,
    \item\label{it2:recons}  $\bigcup_{1\leq i \leq r} \{N_G(K_i)\}$ is a nested family,
    \item\label{it3:recons}  $(K_i \times N_G(K_i)) \subseteq E$ for every $1\leq i \leq r$.
\end{enumerate}
\end{theoremEnd}

\begin{proofApx}  We first prove the forward direction. Assume that $G$ is a trivially perfect 
graph and that $S \subseteq V$ is a maximal clique of $G$. We have that: 
\begin{itemize}
    \item \ref{it1:recons} is trivially true by heredity of trivially perfect graphs.
    \item \ref{it2:recons} holds since otherwise there would be a $P_4$ with $x,y \in S$, $z\in K_i, w \in K_j$ for some $1 \leqslant i < j \leqslant r$ such that $zx, yw \in E, xw, yz \notin E$ ($K_i$ and $K_j$ are chosen such that $N_G(K_i)$ and $N_G(K_i)$ are not comparable w.r.t. inclusion).
    \item \ref{it3:recons} 
    Let $x \in K_i$ be a vertex with a neighbor $z \in S$, $1 \leqslant i \leqslant r$. 
    We will first show that every neighbor $y$ of $x$ in $K_i$ is also a neighbor of $z$. By contradiction, suppose that $yz \notin E$. By maximality of the clique $S$, there exists $w \in S$ non-adjacent to $x$. If $w$ is adjacent to $y$ then the vertices $\{w,z,x,y\}$ induce a $C_4$, else they induce a $P_4$, a contradiction in both cases.
    
    Since $G[K_i]$ is connected, the argument that $z$ is adjacent to the neighbor of $x$ extends to every vertex of $K_i$, $1 \leqslant i \leqslant r$. 
    Applied to each $z \in N(K_i)$, this shows that the set $E(G)$ contains all edges between $K_i$ and $N(K_i)$ for every $1 \leqslant i \leqslant r$.
\end{itemize}

\noindent We now turn our attention to the reverse direction. 
Let $G$ be a graph that verifies the conditions~\ref{it1:recons},~\ref{it2:recons} and~\ref{it3:recons}. 
Since trivially perfect graphs are hereditary, condition~\ref{it1:recons} implies that $G[K_i]$ is trivially perfect so it does not 
contain any $P_4$ nor $C_4$ as an induced subgraph. 
We will now show that $G$ does not contain any obstruction. 
By contradiction, let $W \subseteq V$ be an obstruction in $G$. Notice that 
$|W \cap K_i| \geq 2$ is impossible because, according to \ref{it3:recons}, every vertex of $K_i$ has the same neighborhood outside of $K_i$.
Moreover, $|W \cap S| \geq 3$
 is also impossible because $S$ is a clique. Finally, 
 $W$ cannot intersect three distinct connected components $K_i$ since otherwise it
 would be a claw or disconnected. Thus $W$ intersects $K_i, K_j, i\neq j$ and $|W \cap S| = 2$. According to \ref{it2:recons}, $N_G(K_i) \subseteq N_G(K_j)$ or $N_G(K_j) \subseteq N_G(K_i)$ holds, 
 and we thus conclude that $W$ cannot be an obstruction.
 \end{proofApx}

\section{Kernelization algorithm for {\sc \TPE{}}}
\label{sec:rules}

\begin{figure}[ht]
    \centering
    \includegraphics[scale=1.75]{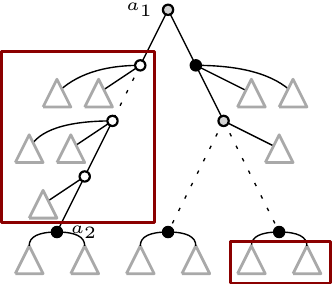}
    \caption{Analysis of a universal clique decomposition of a connected trivially perfect graph. 
    Black vertices represent bags of $A$, gray vertices bags of $A'$ and  
    triangles are connected trivially perfect subgraphs of $G$. The leftmost rectangle is a comb of $G$, the rightmost a trivially perfect module.  Note that any group of triangles rooted at a same bag is a trivially perfect module. \label{fig:size}} 
\end{figure}

We begin this section by providing a high-level description of our kernelization 
algorithm. As mentioned in the introductory section, we use the universal clique decomposition of trivially perfect graphs to bound the number of vertices of a reduced instance. Let us consider a positive instance $(G = (V,E),k)$ of \TPE{}, $F$ a suitable solution and $H = G \triangle F$. 
Denote by $\mathcal{T}=(T,\mathcal{B})$ the 
universal clique decomposition of $H$ as described Definition~\ref{def:ucd}. Since 
$|F| \leqslant k$, we know that at most $2k$ bags of $\mathcal{T}$ may contain affected vertices. Let 
$A$ be the set of such bags, and let $A'$ denote the least common ancestor closure of 
$A$ in forest $T$ (Definition~\ref{def:lca-closure}). 
As we shall see later, the size of 
$A'$ is also linear in $k$ (Lemma~\ref{lemme:borne_lca}). 
The removal of every bag of $A'$ from $T$ will disconnect the forest $T$ into several components (see Figure~\ref{fig:size}). 
Such a connected component $D$ of $T \setminus A'$ may see zero, one or two nodes of $A'$ in the forest $T$ (Lemma~\ref{lem:ucd}). If $D$ has no neighbour in $A'$, the union of all bags of $D$ corresponds to a connected component of $H$ and of $G$, inducing a trivially perfect graph in $G$, and will be eliminated by a reduction rule. We shall see that the union of all components $D_a$ of the second type, seeing a unique bag $a \in A'$ in the forest $T$, corresponds to a trivially perfect module of graph $G$. We use the reductions rules of~\cite{DP18} to shrink such a module to $O(k^2)$ vertices, which boils down to a total $O(k^3)$ vertices since $|A'| = O(k)$. 
Our efforts will be focused on components $D$ seeing two bags $a_1, a_2 \in A'$, one of them being ancestor of the other in forest $T$. We call such a structure $D$ a \emph{comb} (Definition~\ref{def:comb} and Figure~\ref{fig:size}). 
Such combs (the union of their bags) induce, in graph $G$, a trivially perfect subgraph that can be partitioned 
with regard to critical cliques and trivially perfect modules with nice inclusion properties on their 
neighborhoods. 
We provide two distinct reduction rules on these structures. Rule~\ref{rule:peigneTP} reduces the so-called shaft of the comb (intuitively, the path strictly between $a_1$ and $a_2$ in $T$) to length $O(k)$. Rule~\ref{rule:Reduction_dents} reduces the size of the whole comb (the union of its bags) to $O(k^2)$. 
Altogether, the reduced instance cannot contain more 
than $O(k^3)$ vertices. \\

We would like to note that the combs considered in this work are similar to the ones defined by Drange and Pilipczuk~\cite{DP18} and thus named after them. 
However, the two structures are not strictly 
identical, in particular since they were originally defined with respect to a vertex modulator (\ie a packing of obstructions), and 
thus their neighborhood towards the rest of the graph was structured differently.  \\

In the remaining of this section we assume that we are given an instance $(G = (V,E), k)$ of \TPE{}. 

\subsection{Reducing critical cliques and trivially perfect modules}

We first give a classical reduction rule when dealing with modification problems. This rule is safe for any target graph class hereditary and closed under disjoint union. Notice 
that this rule will allow us to reduce connected components of $T \setminus A'$ having 
no neighbor in $A'$. 

\begin{polyrule}
    \label{rule:compTP}
   Let $C \subseteq V$ be a subset of vertices such that $G[C]$ is a trivially perfect connected component of $G$. %
    Remove $C$ from $G$. 
\end{polyrule}

We now give known reduction rules that deal with critical cliques and trivially 
perfect modules.  
The safeness of Rule~~\ref{rule:borneCC-TP} comes from the fact that 
trivially perfect graphs are hereditary and closed under true twin addition combined with 
Lemma~\ref{lem:homogen-compl-here}. The safeness and polynomial-time application 
of Rule~\ref{rule:borneIdp-TP} was proved 
by Drange and Pilipczuk~\cite{DP18}. We would like to mention that while the statement of their 
rule assumes the instance at hand to be reduced by classical \emph{sunflower} rules, this 
is actually not needed to prove the safeness of the rule. Altogether, we have the following.

\begin{polyrule} \label{rule:borneCC-TP}
    Let $K \subseteq V$ be a set of true twins of $G$ such that $|K| > k+1$. Remove $|K|-(k+1)$ arbitrary vertices in $K$ from $G$.
\end{polyrule}

\begin{polyrule} \label{rule:borneIdp-TP}
    Let $M \subseteq V$ be a module of $G$ such that $G[M]$ is trivially perfect and 
    $M$ contains an independent set $I$ of size at least $2k+5$. Remove all vertices 
    of $M \setminus I$ from $G$. 
\end{polyrule}

\begin{theoremEnd}{lemma}[Folklore,\cite{BPP10,DP18}]
\label{lem:simple}
    Rules~\ref{rule:compTP} to~\ref{rule:borneIdp-TP} are safe and can be applied in 
    polynomial time. 
\end{theoremEnd}

Using a structural result on trivially perfect graphs where critical cliques and independent 
sets have bounded size, Drange and Pilipczuk~\cite{DP18} proved the following. 

\begin{lemma}[\cite{DP18}] \label{lem:bornemoduleTP}
Let $(G=(V,E),k)$ be an instance of \TPE{} reduced under Rules~\ref{rule:borneCC-TP} and~\ref{rule:borneIdp-TP}. Then for every module $M \subseteq V$ such that $G[M]$ is trivially 
perfect, $|M| = O(k^2)$. 
\end{lemma}

\subsection{Reducing shafts of combs} 

\begin{figure}[h]
    \centering
    \includegraphics[scale=1.5]{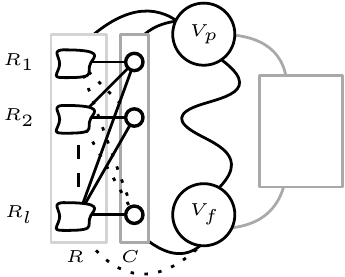}
    \caption{Illustration of a comb, with shaft $C$ and teeth $R$. The edges between $V_p$ and $V_f$ can be anything. Every tooth $R_i$ induces a (possibly disconnected) trivially perfect module. 
    \label{fig:comb}}
\end{figure}

We now consider the main structure of our kernelization algorithm, namely \emph{combs}. 
Recall that such structures are similar to the ones defined by Drange and Pilipczuk~\cite{DP18} 
but not strictly identical. More precisely, the inner part of the structure is the same but not 
their neighborhoods towards the rest of the graph. We however choose to use the same name since 
it is well-suited to illustrate the structure (see Figure~\ref{fig:comb}).

\begin{definition}[Comb]\label{def:comb}
 Let $G=(V,E)$ be a graph and $C ,R  \subseteq V$ be such that $C$ is a clique which can be partitioned into $l$ critical cliques $\{C_1,...,C_l\}$ and $R$ can be partitioned into $l$ non-empty and non-adjacent trivially perfect modules $\{R_1,...,R_l\}$. The pair $P = (C,R)$ is a \emph{comb} if and only if: 
\begin{itemize}
    \item there exist $V_f,V_p \subseteq V(G)\backslash\{C,R\}$, $V_f \neq \emptyset$ such that $\forall x \in C,\ N_{G\backslash (C \cup R)}(x) = V_p \cup V_f$ and $\forall y\in R,\ N_{G\backslash (C \cup R)}(y) = V_p$,
    \item $N_{G[R]}(C_i)= \bigcup_{j=i}^l R_j$ and $N_{G[C]}(R_i)= \bigcup_{j=1}^i C_j$ for $1 \leq i \leq l$.
\end{itemize}
\end{definition}

\noindent Proposition~\ref{prop:unique} states that given a comb $(C,R)$ of graph $G=(V,E)$, the subgraph $G[C \cup R]$ is trivially perfect, and has a universal clique decomposition in which critical cliques $(C_1, \dots, C_l)$ are arranged in a path starting from the root, the \emph{shaft} of the comb, and the decomposition of each \emph{tooth} $R_i$ is attached to $C_i$; see Figure~\ref{fig:comb}.
The length of $(C,R)$ is $l$, the number of critical cliques in $C$. We can observe that 
$N_G[C_l]\subsetneq \dots \subsetneq N_G[C_1]$ and $N_G(R_1) \subsetneq \dots \subsetneq N_G(R_l)$ because 
for $1\leq i \leq l$, $N_G[C_i] = (\bigcup_{j=i}^l R_j) \cup V_p \cup V_f $  and $N_G(R_i) = 
(\bigcup_{j=1}^i C_j) \cup V_p$.

\begin{theoremEnd}[category=rules]{proposition}\label{prop:unique}
Given a comb $(C,R)$ of graph $G=(V,E)$, the subgraph $G[C \cup R]$ is trivially perfect. Moreover the sets $V_p$ and $V_f$, and the ordered partitions $(C_1,\dots, C_l)$ of $C$ and $(R_1,\dots,R_l)$ of $R$ are uniquely determined.
\end{theoremEnd}

\begin{proofApx}
Observe that $V_p = N_{G\setminus C}(R)$, and  $V_f = N_{G\setminus R}(C) \setminus V_p$, thus both sets are uniquely determined.
Since $(C,R)$ is a comb, $C$ is a union of $l$ critical cliques that are totally ordered by the inclusion of their closed neighborhood, $N_G[C_l]\subsetneq \dots \subsetneq N_G[C_1]$. Lastly, for all $i,\ 1 \leq i <l$, $R_i = N(C_i) \setminus N(C_{i+1})$ and $R_l = R \setminus \cup_{i=1}^{l-1} R_i$ by definition of a comb. 
Therefore $G[C \cup R]$ admits a universal clique decomposition, where each $G[R_1],..., G[R_{l-1}]$ admits a universal clique decomposition, and the nodes corresponding to $C_i, 1 \leq i < l-1$ are parents of node $C_{i+1}$ and of the root of the decomposition of $G[R_i]$. The node corresponding to $C_{l-1}$ is connected to the decomposition of $G[R_{l-1}]$ and $G[R_l \cup C_l]$, which admits a universal clique decomposition since $G[R_l]$ is trivially perfect and $C_l$ is a universal clique of $G[R_l \cup C_l]$.
\end{proofApx}

\begin{theoremEnd}[normal,category=rules]{lemma}\label{lem:longueur_peigne}
    Given an instance $(G = (V,E), k)$ of \TPE{} and a comb $(C,R)$ of length $l \geq 2k+2$ of $G$, there is no $k$-edition that affects vertices in $C \cup R$.
\end{theoremEnd}

\begin{proofApx}
Consider a $k$-edition $F$ of $G$  
and $H = G \triangle F$. Denote by $F' \subseteq F$ the subset of pairs from $F$ which 
does not contain any vertex from $C\cup R$ and let $H' = G \triangle F'$. Since $|F| \leq k$ and $(C,R)$ is a comb of length $\geq 2k+2$, there exist $i\neq j \in \{1,\dots , l\}$ such that $C_i, R_i, C_j$ and $R_j$ do not include  affected vertices of $F$. Let us take $c_1 \in C_i,  r_1 \in R_i, c_2 \in C_j $ and $r_2 \in R_j$.

Suppose that $H'$ is not trivially perfect, then there exists an obstruction $W$ of $H'$ such that $A = W \cap (C\cup R) \neq \varnothing$. 
Since pairs of $F'$ do not contain vertices of $C \cup R$, $(C,R)$ is a comb in $H'$ and $|A| = 4$ is impossible since $H'[C\cup R]=G[C \cup R]$ is trivially perfect by Proposition~\ref{prop:unique}. We show that $|A| = 3$ is also impossible. If $|A| = 3$ then the vertex $x \in W \backslash (C\cup R)$ is in the set $V_p$ or $V_f$, otherwise the obstruction $W$ would not be connected.  
We now show that $H'[W]$ contains a claw (as subgraph), a triangle or is not connected. If $x \in V_p$, then by construction $x$ is adjacent to every vertex of the comb and $H'[W]$ would contain a claw. If $x \in V_f$ and $A$ contains at least two vertices in $C$, then these vertices would induce a triangle with $x$.
If $x \in V_f$ and $A$ contains at least two vertices $r',r'' \in R$, then $x$ is not adjacent to any of them (since $V_f$ does not see $R$ in $G$). If $r'$ and $r''$ are not adjacent in $H'$, either the fourth vertex of $W$ sees $r',r''$ and $x$ so $H'[W]$ contains a claw, or $H'[W]$ is disconnected. If $r'$ and $r''$ are adjacent in $H'$, they must belong to a same module $R_i$. Again the fourth vertex of $W$ must either see them both thus forming a triangle, or none of them and $H'[W]$ is disconnected. In any case, $A$ cannot be an obstruction and we conclude that either $|A| = 1$ or $|A| = 2$.
 We shall now construct an obstruction $W' = (W\backslash A) \cup A'$ such that $H'[W]$ and $H'[W']$ are isomorphic and $A' \subseteq \{c_1,r_1,c_2,r_2\}$. We can observe that $W$ must contain a vertex from $V_p$ or $V_f$. \begin{itemize}
    \item If $|A| = 1$, take $x\in A$. If $x \in R$ then let $A' = \{r_1\}$, else let $A' = \{c_1\}$. Since $(C,R)$ is a comb, $H'[W]$ and $H'[W']$ are isomorphic.
    \item If $|A| = 2$, denote by $x$ and $y$ the elements of $A$. If $x,y \in C$, then $H[W]$ contains a triangle. If $x\in C$ and $ y\in R$, in the subcase $xy \in E(H')$ let $A' = \{c_1, r_1\}$ and observe that $c_1r_1 \in E(H')$, hence $H'[W]$ and $H'[W']$ are isomorphic; in the other subcase $xy \notin E(H')$, take $A' = \{c_2, r_1\}$, so $c_2r_1 \notin E(H')$ thus again $H'[W]$ and $H'[W']$ are isomorphic. Eventually consider the last case $x,y \in R$. If $xy \in E(H')$ then $H[W]$ contains a triangle, else $xy \notin E(H')$, so let $A' = \{r_2, r_1\}$ and note that $r_2r_1 \notin E(H')$ thus $H'[W]$ and $H'[W']$ are isomorphic.
\end{itemize}
The set $W'$ is an obstruction of $H'$ and since the vertices in $\{c_1,r_1,c_2,r_2\}$ are not incident to any pair of $F$, $W'$ is also an obstruction of $H$. Therefore $H$ is not trivially perfect, which is a contradiction, concluding the proof of the Lemma.  
\end{proofApx}

\begin{polyrule}\label{rule:peigneTP}
Given a comb $(C,R)$ of length $l \geq 2k+2$ of $G$, remove from $G$ the vertices in $C_i \cup R_i$ for $2k+2 < i \leq l$. 
\end{polyrule}

\begin{theoremEnd}[category=rules]{lemma}
Rule~\ref{rule:peigneTP} is safe.
\end{theoremEnd}
\begin{proofApx}
    Let $(G = (V,E), k)$ be an instance of \TPE{}, $(C,R)$ a comb of $G$ of length at least $2k+2$ and 
    $G'$ the graph obtained from the application of Rule~\ref{rule:peigneTP} on 
    $(C,R)$. We can observe that in $G'$ there is a comb $(C',R')$ where 
    $C' = C_1 \cup \dots \cup C_{2k+2}$ and $R' = R_1 \cup \dots \cup R_{2k+2}$. 
    Let $F$ be a $k$-edition of $G$.  Then by heredity, the graph $G' \triangle F$ 
    is trivially perfect. Now, let $F'$ be a $k$-edition of $G'$ and $H' = G' \triangle F'$. 
    By Lemma~\ref{lem:taille_peigne} $F'$ does not affect any vertex of $C' \cup R'$. 
    We now show that the graph $H = G \triangle F'$ is also trivially perfect. 
    For the sake of contradiction, suppose that there exists an obstruction $W$ in 
    $H$. Since $H'$ is trivially perfect, this  
    obstruction must intersect vertices of $(C \cup R) \backslash (C' \cup R')$. 
    Using the same construction as the one in proof of Lemma~\ref{lem:longueur_peigne} 
    with vertices $c_1 \in C_1, c_2 \in C_2, r_1 \in R_1, r_2 \in R_2$, we can 
    construct in $H'$ an obstruction $W'$ isomorphic to $W$ that intersects 
    the comb $(C,R)$
    only on the vertices $\{c_1,r_1,c_2,r_2\}$, which is impossible since $H'$ 
    is trivially perfect.
\end{proofApx}

\subsection{Breaking the teeth}

\begin{theoremEnd}[normal,category=rules]{lemma} \label{lem:bornedent}
    Let $(G = (V,E),k)$ be a yes-instance of \TPE{}, and $(C,R)$ be a comb of $G$ such that there exist $a,b\in \{1,\dots,l\}$ with $\Sigma_{a \leq i \leq l} |R_i| \geq 2k+1$ and $\Sigma_{b \leq i < a} |R_i| \geq 2k+1$. Then there exists an optimal $k$-edition $F$ of $G$ such that for every $m \in \{1, \dots, b-1\}$, the vertices of $R_m$ are all adjacent to the same vertices of $V(G)\backslash R_m$ in $G \triangle F$, and $F$ contains no pair of vertices of $R_m$.
\end{theoremEnd}

\begin{proofApx}
Let $F$ be an optimal $k$-edition of $G$ and $H = G \triangle F$. There exist $v_2 \in (R_a \cup R_{a+1} \cup \dots \cup R_l)$ and $v_1 \in (R_b \cup R_{b+1} \cup \dots \cup R_{a-1})$
unaffected by $F$. The neighborhood of $v_1$ in $H\backslash R$ must be a clique: indeed, if there exist $x,y \in N_{G\backslash R}(v_1)$ such that $xy \notin E(H)$, then since $N_{G\backslash R}(v_1) \subseteq N_{G\backslash R}(v_2)$ the vertices $\{v_1,x,v_2,y\}$ would induce a $C_4$.  
Let $1 \leqslant m < b$, we will construct an edition $F_m$ such that $|F_m| \leqslant |F|$, $F_m$ contains no pair of vertices included 
in $R_m$ and the vertices of $R_m$ are all adjacent to the same vertices 
in $G \triangle F_m$. Applying this construction iteratively to each $R_m$, $1 \leqslant m < b$ will yield an edition $F^*$ that verifies the desired properties.

Let $S$ be a maximal clique in $H$ that contains $N_{G\backslash R}(v_1)$ and $v_1$, and let $K_1,\dots, K_r$ be the connected components of $H \backslash S$. Observe that $K_1, \ldots , K_r$ respect the conditions~\ref{it1:recons},~\ref{it2:recons} and~\ref{it3:recons} of Lemma~\ref{thm:TP_recons} with $S$. Let $v_m \in R_m$ be a vertex incident to the least number of pairs 
of $F$ with an extremity in $S$. 

Denote by $N$ the set of vertices of $S$ adjacent to $v_m$ in graph $H$. Let $H'$ be the graph constructed from $H \backslash R_m$ and $G[R_m]$ by adding the edges $N \times R_m$, and $F_m$ be the edition such that $H' = G \triangle F_m$. By construction $|F_m| \leq |F|$, we will now show that $H'$ is trivially perfect.

We can observe that $R_m \cap S = \emptyset$ (because $v_1$ is unaffected by $F$ and is non-adjacent with $R_m$ in $G$) and therefore that $S$ is a maximal clique of $H\backslash R_m$. 

By construction of $H'$, $S$ is also a maximal clique of $H'$ and $R_m$ is a connected component of $H_m \setminus S$. Let $K_1',\dots, K_{r'}'$ be the connected components of $(H\backslash R_m) \backslash S$. Sets $K_1',\dots, K_{r'}'$ verify the conditions~\ref{it1:recons},~\ref{it2:recons} and~\ref{it3:recons} of Lemma~\ref{thm:TP_recons} with respect to $S$ in $H\backslash R_m$ and thus also in $H'$. Moreover $H'[S \cup R_m]$ is trivially perfect and $(N_{H'}(R_m) \times R_m) \subseteq E(H')$ by construction. The family $\bigcup_{1\leq i \leq r} \{N_H(K_i)\}$ is nested according to Lemma~\ref{thm:TP_recons}, and, by construction of $H'$, $\bigcup_{1\leq i \leq r'} \{N_{H'}(K_i')\} \subseteq \bigcup_{1\leq i \leq r} \{N_H(K_i)\}$.
We also have that $N \in \bigcup_{1\leq i \leq r} \{N_H(K_i)\}$. Indeed, let $K(v_m)$ the connected component of $H \setminus S$ containing $v_m$, according to condition~\ref{it3:recons} from Lemma~\ref{thm:TP_recons} we have $N_H(K(v_m)) = N_H(v_m) \cap S = N$. Therefore the family $\bigcup_{1\leq i \leq r} \{N_{H'}(K_i')\} \cup \{N\}$ is also nested. By Lemma~\ref{thm:TP_recons} applied on $H'$ and $S$, graph $H'$ is trivially perfect. \\

As mentioned previously, we can apply this construction iteratively to each $R_m$, $1 \leqslant m < b$ and obtain an edition $F^*$ that verifies the desired properties.  
\end{proofApx}

\begin{polyrule} \label{rule:Reduction_dents}
Consider a comb $(C,R)$ of $G$ such that there exist $a,b\in \{1,\dots,l\}$ with $\Sigma_{a \leq i \leq l} |R_i| \geq 2k+1$ and $\Sigma_{b \leq i < a} |R_i| \geq 2k+1$. Then for every $i \in \{1, \dots, b-1\}$, replace $R_i$ by a clique of size $min(|R_i|, k+1)$ with the same neighborhood.
\end{polyrule}

\begin{theoremEnd}[category=rules]{lemma}
Rule~\ref{rule:Reduction_dents} is safe.
\end{theoremEnd}

\begin{proofApx}
Let $(G = (V,E), k)$ be an instance of \TPE{} and $(C,R)$ be a comb of $G$ such that there exist $a,b\in \{1,\dots,l\}$, $\Sigma_{a \leq i \leq l} |R_i| \geq 2k+1$ and $\Sigma_{b \leq i < a} |R_i| \geq 2k+1$. Let $G'$ be the graph obtained after applying Rule~\ref{rule:Reduction_dents} on $(C,R)$. We can observe that in $G'$ there is a comb $(C',R')$ where $C' = C_1 \cup \dots \cup C_l$ and $R' = R_1' \cup \dots \cup R_{b-1}' \cup R_b \cup \dots \cup R_l$ where $R'_i$, $1 \leq i < b$ is the clique of size $\min(|R_i|, k+1)$ in $G'$ that replaced the tooth $R_i$ of $G$.

Let $F$ be an optimal $k$-edition of $G$ and $H = G \triangle F$, let us construct a $k$-edition of $G'$. By Lemma~\ref{lem:bornedent} we can assume that the vertices in each tooth $R_i, 1 \leq i < b$ are all adjacent to the same vertices in $V(H)\backslash R_i$. We can observe that if $|R_i| > k$ for $1\leq i < b$ then $F$ does not affect any vertex of $R_i$ or else $|F| > k$. We construct the set $F'$ by removing the pairs $(x,y)$ of $F$ where $x \in R_i, y \in V(G)$ and adding the pairs $(x',y)$ for all $x' \in R_i'$. We have $|F| = |F'|$ by the previous observation (recall that $|R'_i| = min(|R_i|, k+1)$). We will now show that the graph $H' = G' \triangle F'$ is trivially perfect.

Let $S$ be a maximal clique of $H$ constructed as in the proof of Lemma~\ref{lem:bornedent} by taking $v_1 \in (R_b \cup \dots \cup R_{a-1})$ a vertex unaffected by $F$ and $S$ a maximal clique in $H$ that contains $N_{G\backslash R}(v_1)$ and $v_1$. Let $K_1, \dots, K_r$ be the connected components of $H \backslash S$. We can observe that for $ 1 \leq i < b$ each $R_i$ can intersect several sets $K_j, 1 \leq j \leq r$, and if a component $K_j$ intersects some $R_i$, then $K_j \subseteq R_i$. By Lemma~\ref{thm:TP_recons} the family $\bigcup_{1\leq i \leq r} \{N_H(K_i)\}$ is nested. 

Let $K_1',\ \ldots , K_{r'}'$ be the connected components of $H' \backslash S$. We can observe that conditions~\ref{it1:recons} and~\ref{it3:recons} of Lemma~\ref{thm:TP_recons} are verified for every $K_i',\ 1 \leq i \leq r'$ that does not intersect the teeth that were replaced by Rule~\ref{rule:Reduction_dents}. Indeed, these conditions are verified by each $K_i$ and in this case, the editions are the same in $F$ and $F'$ because there exists $K_i$, $ 1\leq i \leq r$ and $K'_j$, $ 1 \leqslant j \leqslant r'$ such that $K_i = K'_j$ and $N_H(K_i) = N_{H'}(K'_j)$. 
Now, notice that the teeth $R_j', 1\leq j < b$ are clique modules in $H'$, thus the connected components $K_i'$, $1 \leq i \leq r'$ containing these teeth verify conditions~\ref{it1:recons} and~\ref{it3:recons} of Lemma~\ref{thm:TP_recons}. Finally the family $\bigcup_{1\leq i \leq r'} \{N_{H'}(K_i')\}$ is nested because it is equal to the family $\bigcup_{1\leq i \leq r} \{N_H(K_i)\}$ (which is nested by Lemma~\ref{thm:TP_recons}): indeed, for each connected component $K_i$, $1 \leq i \leq r$ that contains vertices replaced by Rule~\ref{rule:Reduction_dents}, there is a connected component $K_j'$, $1 \leq j \leq r'$ such that $N_H(K_i) = N_{H'}(K_j')$. Thus the condition \ref{it2:recons} is verified and $H'$ is trivially perfect. \\

Conversely, let $F'$ be an optimal $k$-edition of $G'$ and $H' = G' \triangle F'$. We construct a $k$-edition of $G$, the arguments being very similar to those above.
By Lemma~\ref{lem:bornedent}  
we can assume that the vertices in the tooth $R_i',\ 1 \leq i < b$ are all adjacent to the same vertices in $V(H')\backslash R_i'$. We can observe that if $|R_i'| > k$ for $1\leq i < b$ then $F'$ does not affect any vertex of $R_i'$ or else $|F'| > k$. We construct the set $F$ by removing the pairs $(x',y)$ of $F'$ where $x' \in R'_i,\ y \in V(G)$ and adding the pairs $(x,y)$ for all $x \in R_i$. We have $|F| = |F'|$ by the previous observation (recall that $|R'_i| = min(|R_i|, k+1)$). We will now show that the graph $H = G \triangle F$ is trivially perfect. 

Let $S$ be a maximal clique of $H$ constructed as in the proof of Lemma~\ref{lem:bornedent}, let $K_1', \dots, K_{r'}'$ be the connected components of $H' \backslash S$ and $K_1,\ \ldots , K_r$ be the connected components of $H \backslash S$. We can observe that conditions~\ref{it1:recons} and~\ref{it3:recons} of Lemma~\ref{thm:TP_recons} are verified for every $K_i, 1 \leq i \leq r$ that does not intersect the teeth that were replaced by Rule~\ref{rule:Reduction_dents}. Indeed, these conditions are verified by each $K_i'$ and in this case, the editions are the same in $F'$ and $F$ because there exist $K_i'$, $ 1\leq i \leq r'$ and $K_j$, $ 1 \leqslant j \leqslant r$ such that $K_i' = K_j'$ and $N_{H'}(K_i') = N_H(K_j)$. Now, notice that the teeth $R_j, 1\leq j < b$ are trivially perfect modules in $H$, thus the connected components $K_i$, $1 \leq i \leq r$ containing these teeth verify conditions~\ref{it1:recons} and~\ref{it3:recons} of Lemma~\ref{thm:TP_recons}. Finally the family $\bigcup_{1\leq i \leq r} \{N_H(K_i)\}$ is nested because it is equal to the family $\bigcup_{1\leq i \leq r'} \{N_{H'}(K_i)\}$ (which is nested by Lemma~\ref{thm:TP_recons}): indeed, for each connected component $K_i'$, $1 \leq i \leq r'$ that contains vertices replaced by Rule~\ref{rule:Reduction_dents}, there is a connected component $K_j$, $1 \leq j \leq r$ such that $N_{H'}(K_i') = N_H(K_j)$. Thus the condition \ref{it2:recons} is verified and $H$ is trivially perfect. 
\end{proofApx}

\begin{theoremEnd}[category=rules]{lemma}
\label{lem:taille_peigne} 
Let $(G = (V,E),k)$ be an instance of \TPE{} such that  Rules~\ref{rule:borneCC-TP} to~\ref{rule:Reduction_dents} are not applicable. Then, for every comb $(C, R)$ of $G$, $| C \cup R| = O(k^2)$.
\end{theoremEnd}

\begin{proofApx}
Consider the decomposition of $C$ in critical cliques $C_1,\dots, C_l$ and the decomposition of $R$ in modules $R_1, \dots, R_l$ (Definition~\ref{def:comb}). 
By Rule~\ref{rule:peigneTP}, the comb has at most $l \leq 2k+2$ teeth. Every critical clique $C_i$ is of size at most $k+1$, by  Rule~\ref{rule:borneCC-TP}. Thus $C$ has $O(k^2)$ vertices. 
Among the modules $R_i$, $1 \leq i \leq l$, at most two are of size larger than $2k$, or else the comb would have been reduced by  Rule~\ref{rule:Reduction_dents} (indeed, if there exist $1 \leq c <b <a \leq l$ with $R_c,R_b$ and $R_a$ of size at least $2k+1$, this rule would reduce $R_c$). The two largest modules $R_i$ are of size $O(k^2)$ according to Lemma~\ref{lem:bornemoduleTP}, and since $G$ is reduced by Rule~\ref{rule:Reduction_dents} 
the other teeth are of size $O(k)$, implying that $R$ is of size $O(k^2)$. 
We conclude that the comb is of size $O(k^2)$. 
\end{proofApx}

\section{Enumerating combs}
\label{sec:haircut}

We now prove that reduction rules involving combs can be applied in polynomial time. To that aim, 
we provide an algorithm that can enumerate all \emph{critical combs} in polynomial time. 
A comb $(C,R)$ is said to be \emph{critical} if $R \cup C \cup V_f$ is not a trivially perfect module and if the comb is inclusion-wise maximal, i.e., no other comb $(C',R')$ satisfies $C \subseteq C'$ and $R \subseteq R'$, with one of the inclusions being strict.

\begin{theoremEnd}[normal, category=haircut]{lemma} \label{lem:find_comb}
Algorithm~\ref{al:comb2} enumerates all critical combs of the input graph, in polynomial time.
\end{theoremEnd}

\begin{proofApx}
    We aim at constructing a binary relation over the set of critical cliques of $G$, corresponding to the fact that there is a comb of $G$ such that the first clique $C'$ is the parent of the second clique $C''$ in the universal clique decomposition tree of the comb. 
    
 A necessary condition is that $N[C''] \subsetneq N[C']$, $R(C',C'') =  N[C']\setminus N[C'']$ is a non-empty, trivially perfect module of graph $G$, and $N(R(C',C'')) \subsetneq N(C'')$. Let us denote this relation, constructed by the first loop of Algorithm~\ref{al:comb2}, by $C' \prec C''$.
 
 We prove that if $C^a, C^b$ and $C^c$ are three distinct critical cliques such that $C^a \prec C^c$ and $C^b \prec C^c$, there is no comb of $G$ in which $C^a$ precedes $C^c$ on the shaft. By contradiction, assume the existence of such a comb $(C,R)$ where $C$ is partitioned into critical cliques $(C_1, \dots, C_i=C^a,C_{i+1}=C^c,\dots, C_p)$ and $R$ is partitioned into trivially perfect non-empty modules $(R_1, \dots, R_i=R(C^a,C^c),R_{i+1},\dots, R_p)$, with sets $V_p$ and $V_f$ as in Definition~\ref{def:comb}.
If $C^b$ intersects some tooth $R_j$ of the comb, $1 \leq j \leq l$, then $R_j$ would see $V_f \subseteq N[C^c] \subseteq N[C^b]$~-- a contradiction.
If $C^b$ intersects $V_f$, then $V_f$ sees $R_{i+1} \subseteq N(C^c) \subseteq N[C^b]$ -- a contradiction.
It remains that $C^b$ is one of the critical cliques of the shaft of the comb, or $C^b \subseteq V_p$. Note that in both cases $N[C^a] \subsetneq N[C^b]$. Indeed if $C^b$ is on the shaft it cannot be below $C^c$ by $N[C^c] \subsetneq N[C^b]$, hence $C^b$ is above  
$C^a$ and therefore $N[C^a] \subsetneq N[C^b]$. If $C^b$ is in $V_p$, it sees the whole comb (by Definition~\ref{def:comb}), and also $V_p$ because $V_p \subseteq N[C^c] \subseteq N[C^b]$. We cannot have $N[C^a] = N[C^b]$ because, $C^a$ and $C^b$ being clique modules, 
$C^a \cup C^b$ would be also be a clique module, contradicting the fact that $C^a,C^b$ are maximal clique modules. Therefore $N[C^a]$ is strictly contained in $N[C^b]$, hence there is some vertex $r^b \in N[C^b] \setminus N[C^a]$. By the fact that $C^b \prec C^c$, $R(C^b,C^c) = N[C^b] \setminus N[C^c]$ is a module. But this comes in contradiction with the fact that $R(C^b,C^c)$ contains $r^b$, who does not see $C^a$, and also contains the vertices of $R_i = R(C^a,C^c)$, who see $C_a$. Hence we cannot have a comb containing $C^a$ and $C^c$ as consecutive critical clique modules in its shaft.

Therefore, if for a critical clique $C^c$ there are two others $C^a \prec C^c$ and $C^b \prec C^c$, we can safely remove these two relations (as in the second loop of Algorithm~\ref{al:comb2}), without destroying any parent-child relation between consecutive critical cliques of some comb. \\
 
At this stage, relation $\prec$ is an oriented forest, and for any comb $(C,R)$ of $G$ with shaft $C=(C_1,\dots,C_l)$ and teeth $R=(R_1,\dots,R_l)$, $C_1 \prec C_2 \prec \dots \prec C_l$. Moreover, for any $1 \leq i \leq l$, we must have $R_i = R(C_i,C_{i+1}) = N[C_i] \setminus N[C_{i+1}]$. \\ 
 
Algorithm~\ref{al:comb2} checks (line~\ref{l:checkComb}) that it only enumerates combs, which are not necessarily critical. 
Let $(C',R')$ be a critical comb. By Property~\ref{prop:unique}, $C'$ and $R'$ have unique partitions $(C'_1,\dots,C'_l)$ and $(R'_1,\dots,R'_l)$ as in Definition~\ref{def:comb}, with corresponding sets $V'_p$ and $V'_f$. Any two consecutive critical cliques $C'_i,C'_{i+1}$ of its shaft satisfy $C'_i \prec C'_{i+1}$ by the first part of the proof. In particular the third loop of Algorithm~\ref{al:comb2} will encounter this path 
$C_1 \prec \dots \prec C_l$, with $C_i = C'_i$ for all $1 \leq i \leq l$. By definition of 
combs, we also have $R_i = R'_i$ for any $i$ \emph{strictly} smaller than $l$. Also, the set $X$ constructed by the algorithm (line~\ref{l:X}) satisfies $X = R'_l \cup V'_f$. Nevertheless at this stage we still need to check that set $R_l$ is correctly constructed by the algorithm.
If each connected component of $G[X]$ induces a trivially perfect module of neighborhood $R_{l-1} \cup C_l$, then $C' \cup R' \cup V'_f$ would induce a trivially perfect module. Indeed the universal clique decomposition of $G[C' \cup R']$ (see Proposition~\ref{prop:unique}) would extend to $C' \cup R' \cup V'_f$, replacing the module $R'_l$ by $X$, contradicting the fact that $(C',R')$ is critical. 

Hence at least one component of $G[X]$ is not added to $R_l$ by the algorithm.  
On the other hand, each connected component of $G[R'_l]$ induces a trivially perfect module (since $R'_l$ is itself a trivially perfect module) whose neighborhood is $C \cup V_p = C' \cup V'_p$, implying that $R'_l \subseteq R_l$. By maximality of the comb $(C',R')$, this inclusion cannot be strict. The only remaining possibility is that $R_l = R'_l$, so the comb $(C',R')$ is enumerated by the algorithm, proving the combinatorial part of the lemma. \\

The algorithm is clearly polynomial. A more careful analysis shows that it can implemented to run in $O(n^2m)$ time. Basically, the complexity is given by the first and last \textbf{forall} loops. Both have $O(n^2)$ iterations: there are at most $n$ critical cliques, and since relation $\prec$ is a forest there are $O(n^2)$ paths. Each iteration can run in linear time, in particular, testing that a pair $(C,R)$ is a comb can be performed in time $O(n+m)$.
\end{proofApx}

\begin{figure}[ht]
\begin{center}
\small
\begin{minipage}{.86\linewidth}
\begin{algorithm}[H]
\caption{\small Enumeration of critical combs}
\label{al:comb2}

\label{l:start}
\ForAll{pairs of critical cliques $C',C''$ s.t. $N[C''] \subsetneq N[C']$}
{   
    $R(C',C'') = N_G[C'] \backslash N_G[C'']$\;
    \If{$N(R(C',C'')) \subsetneq N(C'')$ and $R(C',C'')$ is a non-empty, trivially perfect module}
    {
        set $C' \prec C'$\;
    }
}
\ForAll{critical cliques $C'$ having at least two predecessors w.r.t relation $\prec$}
{
    suppress all relations $C'' \prec C'$\; \label{l:delpere}
}
\ForAll{paths $C_1 \prec \dots \prec C_{l-1} \prec C_l$}
    {
    \label{l:forComb}
    Let $R_i = N(C_{i+1}) \setminus N(C_i)$, $\forall 1 \leq i < l$\;
    Let $X = N(C_l) \setminus N(R_{l-1})$\;\label{l:X}
    Let $R_l$ be the union of connected components of $G[X]$ inducing a trivially perfect module whose neighbourhood is $C_l \cup N(R_{l-1})$\label{l:rlconstr}\;
    Let $V_p = X \backslash R_l$\;
    \If{$(C,R)$ is a comb with partitions $C=(C_1,\dots, C_l)$ and $R=(R_1,\dots,R_l)$}
    {
        \label{l:checkComb}
       add $(C,R)$ to the list of combs\;
    }
}
\end{algorithm}
\end{minipage}
\end{center}
\end{figure}

\begin{theoremEnd}[category=polydetect]{lemma} \label{lem:regle_temps_poly}
    Given an instance $(G = (V,E), k)$ of \TPE{}, Rules~\ref{rule:peigneTP} and~\ref{rule:Reduction_dents} can be exhaustively applied in polynomial time.
\end{theoremEnd}

\begin{proofApx}
    Given a comb with its decomposition in cliques and modules, one can determine if Rules~\ref{rule:peigneTP} and~\ref{rule:Reduction_dents} are applicable in linear time by checking the length of the comb and the number of vertices in its teeth, and apply these Rules in linear time. We saw in Lemma~\ref{lem:find_comb} that the critical combs of a given graph can be enumerated in polynomial time and since there is a polynomial number of  combs in a graph, Rules~\ref{rule:peigneTP} and ~\ref{rule:Reduction_dents} can be exhaustively applied on a graph in polynomial time, implying Lemma~\ref{lem:regle_temps_poly}. 
\end{proofApx}

\section{Bounding the size of a reduced instance}
\label{sec:size}

We now prove thoroughly that any reduced yes-instance of \TPE{} contains $O(k^3)$ vertices. 
To that end, we need the following definition and result. 

\begin{definition}[LCA-closure \cite{FLM+12}] \label{def:lca-closure}
Let $T = (V,E)$ be a rooted tree and $A \subseteq V(T)$. The \emph{lowest common ancestor-closure 
(LCA-closure)} $A'$ of $A$ is obtained as follows. Initially, set $A' = A$. Then, as long as there 
exist $x,y \in A'$ whose least common ancestor $w$ is not in $A'$, add $w$ to $A'$. 
The LCA-closure of $A$ is the last set $A'$ obtained using this process. 
\end{definition}

\begin{lemma}[\cite{FLM+12}] \label{lemme:borne_lca}
Let $T = (V,E)$ be a rooted tree, $A \subseteq V(T)$ and $A' = \mbox{LCA-closure}(A)$. Then 
$|A'| \leqslant 2 \cdot |A|$ and for every connected component $C$ of $T \setminus A'$, $|N_T(C)| \leqslant 2$. 
\end{lemma}

\begin{theoremEnd}[normal,category=size]{theorem} \label{thm:TPkernel_edit}
\TPE{} admits a kernel with $O(k^3)$ vertices. 
\end{theoremEnd}

\begin{proofApx}
Let $(G=(V,E),k)$ be a reduced yes-instance of \TPE{} and $F$ a $k$-edition of $G$. Let $H=G\triangle F$ and $\T = (T, \B)$ the universal clique decomposition of $H$. The graph $G$ is not necessarily connected, thus $T$ is a forest. Let $A$ be the set of nodes $t\in V(T)$ such that the bag $B_t$ contains a vertex affected by $F$. Since $|F| \leq k$, we have $|A| \leq 2k$. Let $A' \subseteq V(T)$ be the set containing the nodes of $\mbox{LCA-closure}(A)$ and the root of each connected component of $T$ (in case the closure does not contain them). 
According to Lemma~\ref{lemme:borne_lca} and Rule~\ref{rule:compTP} which implies that there are at most $2k$ connected components in $G$ and thus $2k$ roots, we have $|A'| \leq 6k$. 

\begin{figure}[ht]
    \centering
    \includegraphics[scale=1.65]{figures/sketch-size-connected}~~~~~~~
    \includegraphics[scale=1.5]{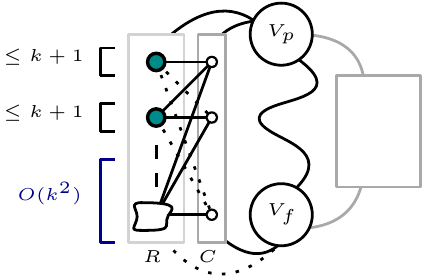}~~
    \caption{(Left) universal clique decomposition of a connected component of $H$. (Right) shape of a reduced comb. \label{fi:subcomb}}
\end{figure}

Let $D$ be a connected component of $T \setminus A'$. We can observe that, by construction of $A'$ (which for every pair of nodes, contains also the smallest common ancestor in $T$), only three case are possibles (see Figure~\ref{fi:subcomb}):
\begin{itemize}
    \item $N_T(D) = \emptyset$ ($D$ is a connected component of $T$).
    \item $N_T(D) = \{a\}$ ($D$ is a subtree of $T$ whose parent is $a \in A'$). 
    \item $N_T(D) = \{a_1,a_2\}$ with one of the nodes $a_1,a_2 \in A'$ being an ancestor of the other in $T$.
\end{itemize}
We will say that these connected components are respectively of type $0$, $1$ or $2$. For $D \subseteq V(T)$, we note $W(D) = \bigcup_{t\in D}B_t$ the set of vertices of $G$ corresponding to bags of $D$. 

There is no connected component of type 0 or else $W(D)$ would be a connected component of $G$ inducing a trivially perfect graph. 
Rule~\ref{rule:compTP} would have been applied to this component, contradicting the fact that $G$ is a reduced instance.

Now consider the set of type 1 components $D_1, D_2,\dots , D_r$ of $T \setminus A'$ attached in $T$ to the same node $a \in A'$. We show that $W_a = W(D_1) \cup W(D_2) \cup \dots \cup W(D_r)$ is a trivially perfect module of $G$. In the graph $H$, $W_a$ is by construction a module of the decomposition. Since no vertex of $W_a$ was affected by the edition $F$, $W_a$ is also a module of $G$, trivially perfect by heredity. 
By Lemma~\ref{lem:bornemoduleTP}, we have $|W_a| = O(k^2)$. There are at most $|A'| \leqslant 6k$ such sets $W_a$, thus the set of vertices of $G$ in bags of type 1 components is of size $O(k^3)$.

Now consider the type 2 connected components $D$ of $T \setminus A'$ which have two neighbor in $T$. 
Let $a_1$ and $a_2$ be these neighbors, one being the ancestor of the other, say $a_1$ is the ancestor of $a_2$. Let $t_1,\dots, t_l$ be the nodes of the tree on the path from $a_1$ to $a_2$, in this order. The component $D$ can be seen as a comb 
of shaft $(B_{t_1}, \dots, B_{t_l})$. 
More precisely, by construction of the universal clique decomposition, $W(D)$ can be 
partitioned into a comb $(C,R)$ of $H$: the critical clique decomposition of $C$ is $(C_1=B_{t_1},\dots, C_l=B_{t_l})$, and each $R_i$ corresponds to the union of bags of the subtrees rooted at $t_i$ which do not contain $t_{i+1}$, for $1 \leqslant i<l$, and to the union of bags of the subtrees rooted at $t_l$ which do not contain $a_2$, for $i=l$. Since $(C,R)$ was not affected by $F$, it is also a comb of $G$. Thus for each type 2 component $D$, $W(D)$ contains $O(k^2)$ vertices by Lemma~\ref{lem:taille_peigne}.
Since $T$ is a forest, it can contain at most $|A'| - 1 \leqslant 6k - 1$ such components in $T \setminus A$. Therefore the set of bags containing type 2 connected components of $T \setminus A$ contains $O(k^3)$ vertices. \\ 
It remains to bound the set of vertices of $G$ which are in bags of $A'$. The vertices corresponding to nodes of $A' \backslash A$ are critical cliques of $G$, and are hence of size at most $k+1$ by Rule~\ref{rule:borneCC-TP}. Thus the set of vertices in bags of $A' \backslash A$ is of size $O(k^2)$. The vertices corresponding to nodes of $A$ are critical cliques in $H$ but not necessarily of $G$. Let $B_a$ be a bag corresponding to a node $a \in A$. We will show that $B_a$ is covered by at most $2k+1$ critical cliques of $G$, which by Rule~\ref{rule:borneCC-TP} will imply that $B_a$ contains $O(k^2)$ vertices of $G$, and thus the set of vertices in bags of $A'$ is of size $O(k^3)$.

To see this, observe that $B_a$ is a critical clique of $H$, and that $G$ is obtained from $H$ by editing at most $k$ pairs of vertices. A result from~\cite{PMS09} claims that, starting from a graph $H$ and editing an edge, we add at most two critical cliques. The same arguments allow to claim that if $B$ is a set of vertices covered by at most $p$ critical cliques in $H$, and if $H'$ is obtained by editing a pair of vertices $x,y$ of $H$, then $p+2$ critical cliques are enough to cover $B$ in $H'$. 
To be complete, we now show this claim. Let $C_1, C_2, \dots, C_p, \dots, C_q$ be the critical cliques of $H$, suppose that $B$ is covered by the first $p$ cliques $C_1,\dots, C_p$. For each $i,\ 1 \leq i \leq q$, the set $C''_i = C_i \setminus \{x,y\}$ is a clique module (not necessarily maximal) of $H'$. In particular, each $C''_i$ is contained in a critical clique $C'_i$ of $H$ (the $C'_i$ are not necessarily distinct). Let $C'(x)$ and $C'(y)$ be the critical cliques of $H'$ containing respectively $x$ and $y$. Clearly, the critical cliques $C'_1, \dots, C'_p, C'(x)$ and $C'(y)$ of $H'$ cover the vertices of $B$, showing our claim. By applying this argument $k$ times (one for each pair of $F$) to the bag $B_a$, which was a critical clique of $H$, we conclude that it is covered by at most $2k+1$ critical cliques of $G$. Thus $|B_a| = O(k^2)$ by Rule~\ref{rule:borneCC-TP}.

We conclude that $|V(G)| = O(k^3)$. Finally, we claim that a reduced instance can be computed in polynomial time. Indeed, Lemma~\ref{lem:simple} states that it is possible to reduce exhaustively a graph under Rules~\ref{rule:borneCC-TP} to~\ref{rule:borneIdp-TP}. Once this is done, it remains to apply exhaustively Rules~\ref{rule:peigneTP} and~\ref{rule:Reduction_dents} which is ensured by Lemma~\ref{lem:regle_temps_poly}.
\end{proofApx}

\section{Kernels for trivially perfect completion/deletion}
\label{sec:variants}

In this section we show that the rules used for \TPE{} are safe for \TPC{} and \TPD{}. First Rules~\ref{rule:compTP},~\ref{rule:borneCC-TP} and \ref{rule:borneIdp-TP} are safe for both problems. Indeed, the safeness of Rule~\ref{rule:borneCC-TP} directly follows from  Lemma~\ref{lem:homogen-compl-here} and Rule~\ref{rule:borneIdp-TP} was shown safe in \cite{DP18}. 
We will now argue that Rules~\ref{rule:peigneTP} and~\ref{rule:Reduction_dents} are also safe. Lemma~\ref{lem:longueur_peigne} states that no trivially perfect edition for an instance $(G = (V,E),k)$ of \TPE{} affects a comb of $G$ of length $\geq 2k + 2$.  This is also true when allowing only edge addition or edge deletion, implying the safeness of Rule~\ref{rule:peigneTP} in both cases. In the proof of Lemma~\ref{lem:bornedent}, for a trivially perfect edition $F$ we construct another edition $F' \subseteq F$. In case $F$ consists only of edge additions or deletions, it is also the case for $F'$, thus Lemma~\ref{lem:bornedent} holds for \TPC{} and \TPD{} and Rule~\ref{rule:Reduction_dents} is safe for these problems. \\

The proof for the size of the kernel is the same as the proof of Theorem~\ref{thm:TPkernel_edit}. Altogether, we obtain the following result. 

\begin{theoremEnd}[category=size]{theorem} \label{thm:TPkernel_cd}
\TPC{} and \TPD{} admits a kernel with $O(k^3)$ vertices. 
\end{theoremEnd}

\section{Conclusion}
We have provided a kernelization algorithm for problem \TPE{}, producing a cubic size kernel,
hence improving upon the $O(k^7)$-size kernel of~\cite{DP18}. The techniques extend to the deletion and completion versions of the problem, within the same bounds. 
A natural question is whether the size of the kernel for \TPE{} can still be reduced -- note that for \TPC{}, Bathie et al.~\cite{BBP21} and Cao and Ke~\cite{CK21} claim a quadratic kernel.  
Some ideas used in this work remind of very similar techniques applied to kernelization problems for edge editing towards classes of graphs $\mathcal{G}$ having a tree-like decomposition. The simplest case -- like here or for the class of so-called 3-leaf power graphs, see~\cite{BPP10} -- is when the vertices of the graph can be partitioned into bags inducing modules, and these bags can be structured as nodes of a forest $T$, with specific adjacency rules. If an arbitrary graph $G$ can be turned into a graph of class $\mathcal{G}$ by editing at most $k$ pairs of vertices, the edited pairs are in some set $A$ of at most $2k$ bags. Again by taking the lowest common ancestor closure $A'$ of $A$, set $A'$ is of size $O(k)$ and its removal from forest $T$ will produce some chunks attached in $T$ to 0, 1 or 2 nodes of $A'$ (e.g., in~\cite{BPP10}, the authors speak of 1 and 2-branches, playing similar roles to modules and combs in this article). Kernelization algorithms can be obtained if we are able to reduce the bags themselves as well as the chunks, which hopefully have good structural properties.
It is natural to wonder how general are these techniques, especially on subclasses of chordal graphs.

\bibliographystyle{plain}
\bibliography{mybib}

\end{document}